\documentclass[fleqn,10pt]{wlscirep}

\usepackage{graphicx}

\usepackage{array}

\makeatletter
\newcommand{\thickhline}{%
    \noalign {\ifnum 0=`}\fi \hrule height 1pt
    \futurelet \reserved@a \@xhline
}
\newcolumntype{"}{@{\hskip\tabcolsep\vrule width 1pt\hskip\tabcolsep}}
\makeatother

\title{Multiplex model of mental lexicon reveals explosive learning in humans}

\author[1,2,*]{Massimo Stella}
\author[3,]{Nicole M. Beckage}
\author[1,]{Markus Brede}
\author[2,4]{Manlio De Domenico}
\affil[1]{Institute for Complex Systems Simulation, University of Southampton, UK}
\affil[2]{Fondazione Bruno Kessler, Italy}
\affil[3]{Department of Electrical Engineering and Computer Science, University of Kansas, USA}
\affil[4]{School of Computer Science and Mathematics, Universitat Rovira i Virgili, Spain}

\affil[*]{Corresponding author: massimo.stella@inbox.com}

\begin{abstract}

Word similarities affect language acquisition and use in a multi-relational way barely accounted for in the literature. We propose a multiplex network representation of this mental lexicon of word similarities as a natural framework for investigating large-scale cognitive patterns.
Our representation accounts for semantic, taxonomic, and phonological interactions and it identifies a cluster of words which are used with greater frequency, are identified, memorised, and learned more easily, and have more meanings than expected at random. This cluster emerges around age 7 through an explosive transition not reproduced by null models. We relate this explosive emergence to polysemy -- redundancy in word meanings. Results indicate that the word cluster acts as a core for the lexicon, increasing both lexical navigability and robustness to linguistic degradation. Our findings provide quantitative confirmation of existing conjectures about core structure in the mental lexicon and the importance of integrating multi-relational word-word interactions in psycholinguistic frameworks.

\end{abstract}

\begin{document}

\flushbottom
\maketitle

\thispagestyle{empty}

\section*{}

Investigating relationships between words offers insights into both the structure of language and the influence of cognition on linguistic tasks \cite{karuza2016local,beckage2015cognition}. As a result, cognitive network science is rapidly emerging at the interface between network theory, statistical mechanics, and cognitive science \cite{de2016large,baronchelli2013,beckage2015cognition,karuza2016local}. The field is influenced by the seminal work of Collins and Quillian\cite{collins1969retrieval}, who assumed that concepts in the human mind are cognitive units, each representable as a node linked to associated elements. These connections represent a complex cognitive system known as the mental lexicon \cite{aitchison2012}. Extensive empirical research has shown that relationships in the lexicon can be modelled as a network of mental pathways influencing both how linguistic information is acquired \cite{storkel2002restructuring,beckage2015cognition,vitevitch2012complex,casas2016polysemy,carlson2014children,hills2009longitudinal}, stored \cite{aitchison2012,storkel2002restructuring,vitevitch2008can,de2016large}, and retrieved \cite{collins1975spreading,vitevitch2012complex,de2016large,i2001small}. 

The cognitive role of quantifying lexical navigability as distances in a network finds empirical
support in several experiments related to word identification and retrieval tasks
\cite{collins1975spreading,collins1969retrieval,dehaene1998imaging,meyer1971facilitation}. For instance, Collins and Loftus\cite{collins1975spreading} showed a correlation between network topology of semantic networks and word processing times: words farther apart in the network require longer identification times, thus indicating higher cognitive effort. More recently, the structural organisation of mental pathways among words was analysed in several large-scale investigations, considering similarity of words in terms of their semantic meaning \cite{sigman2002,de2016large,dorogovtsev2001language}, their phonology \cite{vitevitch2008can,vitevitch2012complex,siew2013community,stella2015patterns,stella2016investigating}, or their taxonomy \cite{i2001small,picard2009hierarchies,liu2014empirical}. Remarkably, all these networks, based on different definitions of relationships between words, were found to be highly navigable: words were found to be clustered with each other and separated by small network distances (sometimes called small-world networks \cite{watts1998collective}). This may suggest a universal structure of language organisation related to minimising cognitive load while maximising navigability of words \cite{sole2015ambiguity,baronchelli2013,beckage2011small,beckage2015cognition}.

The above studies, however, have not yet attempted to use multi-relational information for characterising and quantifying the mental lexicon, instead focusing on only one relationship at a time \cite{collins1975spreading,beckage2011small,hills2009longitudinal,carlson2014children,vitevitch2008can,sigman2002,de2016large,dorogovtsev2001language}. Some researchers have considered the aggregation of several of these relationships into single-layer networks \cite{sigman2002} and others have considered multi-relational models but only to capture the syntactic structure of language \cite{liu2014empirical}. The above approaches offer only limited insight into the cognitive complexity that allow individuals to use language \cite{aitchison2012} with diversity and ease.

More information about the lexical structure can indeed be obtained by accounting, simultaneously, for multiple types of word-word interactions. A natural and suitable framework for this purpose are multilayer networks \cite{de2013mathematical,kivela2014multilayer,boccaletti2014structure,de2016physics,battiston2017new}. Multilayer networks simultaneously encode multiple types of interaction among units of a complex networked system. Therefore, they can be used to extract information about linguistic structures beyond information available from single-layer network analysis \cite{stella2016multiplex}. The usefulness of multiplex representations has recently been shown for diverse applications including the human brain\cite{de2017multilayerbrain,bassett2017network}, social network analysis\cite{szell2010multirelational,mucha2010community,de2015identifying}, transportation\cite{cardillo2012emergence,de2014navigability} and ecology \cite{stella2016parasite,pilosof2017multilayer}.

Here, on an unprecedented scale and from a multi-relational perspective, we investigate the semantics, phonology, and taxonomy of the English lexicon as a model of distinct layers of a multiplex network (see Fig.~1). We study the evolution of multiplex connectivity over the developmental period from early childhood (2 years of age) to adulthood (21 years of age) also through the use of word attributes (e.g. word frequency, length, etc.) influencing lexical acquisition \cite{aitchison2012,kuperman2012age,brysbaert2014concreteness}.

The proposed multiplex representation provides a powerful framework for the analysis of the mental lexicon, allowing for the capture of sudden structural changes that can not be identified by traditional methods. More specifically, when modelling lexical growth, we observe an explosive emergence of a cluster of words in the lexicon around the age of 7 years, which is not observed in single-layer network analyses. We show that this cluster is beneficial from a cognitive perspective, as its sudden appearance facilitates word processing across connected network pathways across all lexicon layers. This boost to cognitive processing also enhances the resilience of the lexicon network when individual words become progressively inaccessible, such as what may happen in cognitive disorders like anomia\cite{laine2013anomia}. These findings represent the first quantitative confirmation and interpretation of previous conjectures about the presence and cognitive impact of a core in the human mental lexicon
\cite{barsalou2008grounded,solonchak2015lexicon,picard2009hierarchies,aitchison2012}.

\section*{Results}
\subsection*{Structure of the Multiplex Lexical Representation}

\begin{figure}[ht]
\centering
\includegraphics[width=15cm]{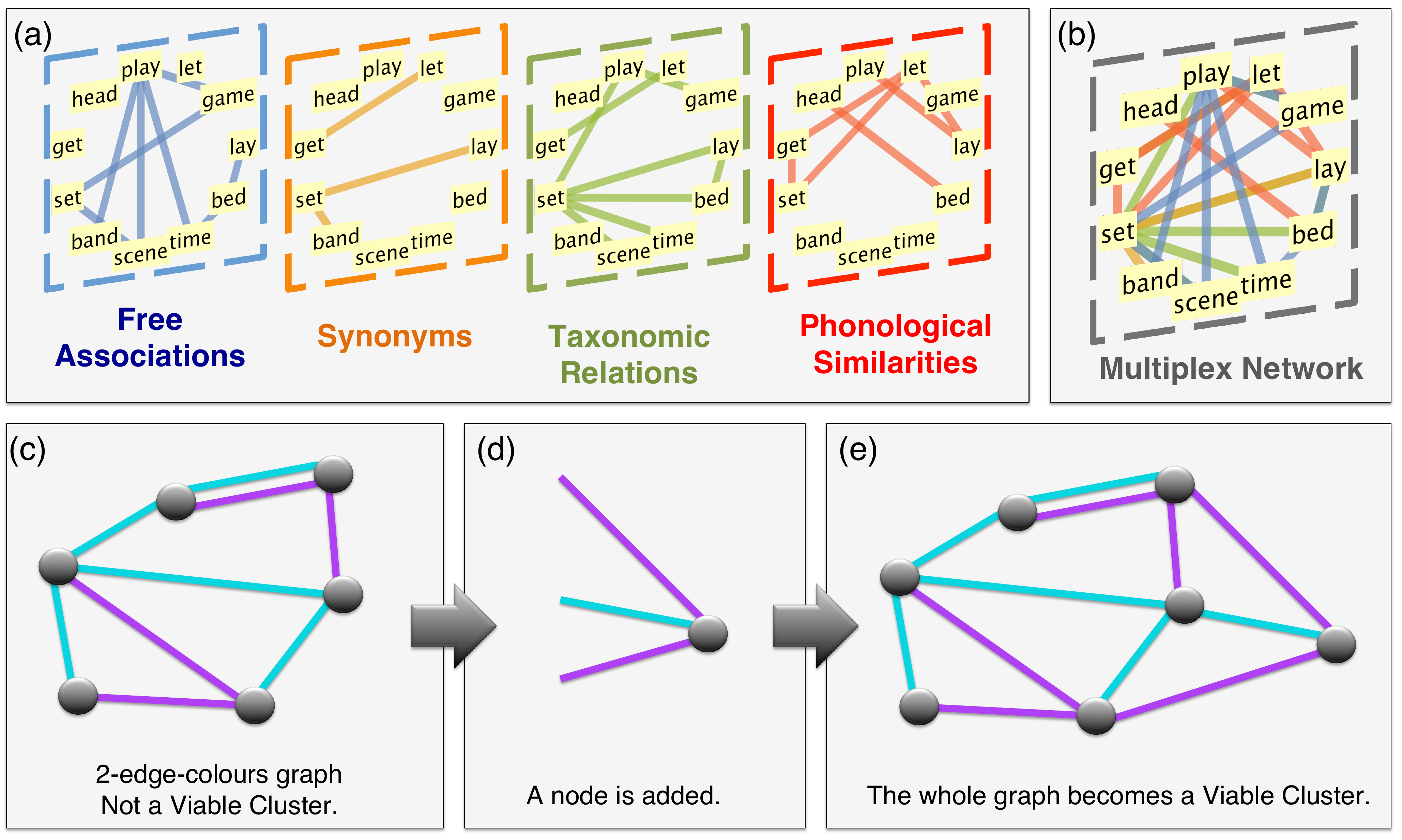}
\caption{(a): Visual representation of a subset of the multiplex lexical representation (MLR) for adults with $N=8531$ words and four types of word relationships forming individual layers: free associations, synonyms, taxonomic relations, and phonological similarities. (b) Multiplex visualisation as an edge-coloured network. (c) Using only purple links does not allow navigation of the whole network. Therefore the network is not a viable cluster. Notice, however, that the two nodes with overlapping links constitute the smallest possible viable cluster in a simple graph (which we refer to as ``trivial'' in the main text). (d-e) The appropriate addition of one node and three coloured links makes the resulting graph a viable cluster, with paths between all nodes using either only cyan or only purple colours.}
\label{fig1}
\end{figure}

Our multilayer lexical representation (MLR) of words in the mind is a multiplex 
network\cite{wasserman1994social,kivela2014multilayer,baxter2016unified,de2016physics} made of $N=8531$ words and four layers. Each layer encodes a distinct type of word-word interaction (cf.\ Fig.\,\ref{fig1} (a)): (i) empirical free associations \cite{coltheart1981mrc}, (ii) synonyms \cite{miller1995wordnet}, (iii) taxonomic relations \cite{miller1995wordnet}, and (iv) phonological similarities \cite{vitevitch2008can}. As shown in Fig.\,\ref{fig1} (b), different relationships can connect words that would otherwise be disconnected in some single-layer representations. We considered these relationships with the aim of building a representation accounting for different types of semantic association, either from dictionaries (i.e.\ synonyms and taxonomic relations) or from empirical experiments (i.e.\ free associations). We also include sound similarities (i.e.\ phonological similarities) as they are involved in lexical retrieval
\cite{vitevitch2008can,vitevitch2012complex}. This set of relationships represents a first approximation to the multi-relational structure of the mental lexicon. Compared to previous work on multiplex modelling of language development \cite{stella2016multiplex}, our multiplex representation is enriched with node-level attributes related to cognition and language: (i) age of acquisition ratings \cite{kuperman2012age}, (ii) concreteness ratings \cite{brysbaert2014concreteness}, (iii) identification times in lexical decision tasks \cite{keuleers2012british}, (iv) frequency of word occurrence in Open Subtitles \cite{barbaresi2014language}, (v) polysemy scores, i.e.\ the number of definitions of a word in WordNet, used to approximate polysemy in computational linguistics \cite{sigman2002,casas2016polysemy} (cf.\ Methods and SI Sect.\ 12) and (vi) word length \cite{kuperman2012age}. The analysis of structural reducibility of our multiplex model (cf.\ SI Sect.\ 2) quantifies the redundancy of the network representation \cite{de2015structural}. Results suggest that no layers should be aggregated, as each network layer contributes uniquely to the structure of the multiplex representation, confirming the suitability of the multiplex framework for further investigation.

As already discussed, investigating navigation on linguistic networks has proved insightful \cite{collins1969retrieval,collins1975spreading,sigman2002}. Hence we focus on analysing the navigability of our multiplex network \cite{de2014navigability}, identifying word clusters that are fully navigable on every layer, i.e.\ where any word can be reached from any other word on every layer when considered in isolation. An example is reported in Fig.\,\ref{fig1} for a representative multiplex network with two layers. In network theory, these connected subgraphs are also called viable clusters \cite{baxter2016unified} (cf.\ Methods). Notice that the largest viable cluster of a single-layer network coincides with its largest connected component \cite{newman2011structure}, i.e. the largest set of nodes that can all be reached from each other within one layer. In multiplex networks the two concepts are distinct, as viable clusters are required to be connected on \textit{every layer when considered individually}. Removing this constraint of connectedness on every layer leads to the more general definition of multi-layer connected components\cite{de2014navigability}, i.e.\ the largest set of nodes all connected to each other when jumps across layers are allowed. Fig.\,\ref{fig1} (c-e) conveys the idea that the emergence of viable clusters can be due to the addition of particular links in the network. 

Our multiplex model contains a single non-trivial (i.e.\ with more than two nodes) viable cluster composed of 1173 words, about 13.8\% of the network size. In the following we refer to this cluster as the largest viable cluster (LVC). For easier reference, we indicate words in the empirical LVC as ``LVC-in words'' and words outside of the empirical LVC as ``LVC-out words''. Reshuffling network links while preserving word degrees leads to configuration model-layers \cite{newman2011structure} that still display non-trivial LVCs (cf.\ LVC Rew.\ in Tab.\ 1). Further, on average $98.1 \pm 0.1 \%$ of LVC-in words persist in the viable cluster after rewiring 5\% of all the intra-layer links at random. We conclude that the LVC does not break but rather persists also in the case of potentially missing or erroneous links in the network dataset (e.g.\ spurious free associations or mistakes in phonological transcriptions).

In order to further test correlations between network structure and word labels, we also consider a full reshuffling null model (see SI Sect.\ 4), in which word labels are reshuffled independently on every layer and thus word identification across layers is not preserved. Hence, full reshuffling destroys inter-layer correlations but preserves network topology. Fully reshuffled multiplex networks did not display any non-trivial viable clusters, emphasizing the important role of inter-layer relationships for the presence of the LVC in the empirical data.

In the next section we analyse the evolution of the LVC during language learning over a time period of more than 15 years. We demonstrate the existence of an explosive phase transition \cite{baxter2016unified} in the emergence of the LVC and explore the significance of this transition from the perspective of cognitive development.

\subsection*{Emergence of the Largest Viable Cluster}

\begin{figure}[h]
\centering
\includegraphics[width=15cm]{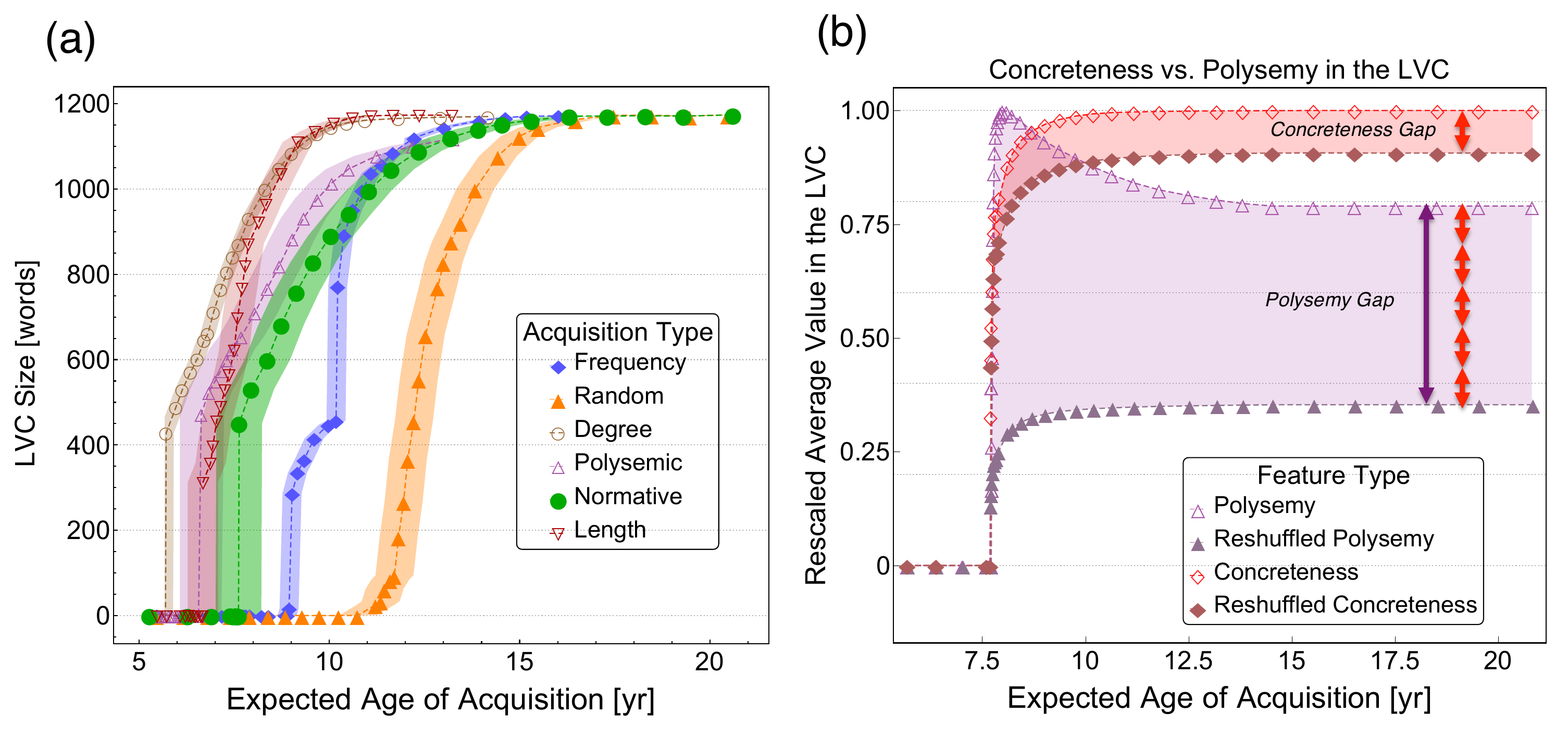}
\caption{(a): Evolution of the size of the LVC when words are acquired in ascending order based on: age of acquisition (green dots), frequency (blue diamonds), polysemy scores (purple triangles), multidegree in the multiplex (brown circles), word length (red upside-down triangles) and at random (orange triangles). The LVC emerges with an explosive transition at $7.7 \pm 0.6$ years in normative acquisition. Areas represent standard deviations considering randomisations of smeared age of acquisition or ties in the rankings. For further details on the concreteness model see SI Sect.\ 15. (b): Comparison of average linguistic features for words in the LVC with normative acquisition in the empirical data and for a partial reshuffling null model with reshuffled node attributes. The curves are rescaled from 0 to 1 by their empirical maximum value and they represent averages over 200 iterations. Error margins are approximately the same size as the dots. Reshuffling node attributes results in an LVC with both reduced concreteness and polysemy scores. We note significant gaps between the empirical and randomised data. The observed gap in polysemy scores is almost 5 times larger than for concreteness values.}
\label{fig2}
\end{figure}

To study the emergence of the LVC during cognitive development, we simulate probabilistic normative word orderings by smearing the age of acquisition dataset\cite{kuperman2012age}. We refer to these orderings as normative acquisition. Smearing allows us to account for the variance in age of acquisition across individuals by introducing a probabilistic interpretation of these orderings (see Methods). We compare the trajectories of normative acquisition against five null models: (i) random word learning (i.e.\ words are acquired at random), (ii) frequency word learning (i.e.\ higher frequency words are acquired earlier), (iii) polysemy-scores word learning (i.e.\ words with a higher count of context-dependent meanings are learned earlier) and (iv) multidegree word learning (i.e.\ words with more connections --across all layers-- are learned earlier) and (v) word length learning (i.e.\ shorter words are learned earlier). We investigate if modelling the development of the mental lexicon as growth of the empirical multiplex representation according to a given learning scheme matches the explosive transition observed in normative learning. Results are reported in Fig.~2 (a).

Normative acquisition indicates a sudden emergence of the LVC around age $7.7 \pm 0.6$ years, almost four years earlier than expected if learning words at random. Further analysis reveals two distinct patterns. Firstly, this sudden appearance is robust to fluctuations in word rankings in the age of acquisition ratings (AoA): in all simulations based on AoA reports, after roughly 2500 words have been acquired, an LVC with at least 260 words suddenly appears with the addition of just a single word to the lexicon. Secondly, the average magnitude of this explosive change is $\Delta L_{AoA}=(420\pm 50)$ words. These patterns suggest an explosive phase transition\cite{souza2015anomalous,grassberger2015percolation,baxter2016unified} in the structural development of the mental lexicon. To the best of our knowledge, this work is the first detection of an explosive change in lexicon structure in cognitive network science during vocabulary growth.
 
Explosive behaviour in the emergence of the LVC is not observed in the random acquisition null model (see Methods and SI Sect.\ 7-11), with only a few cases ($\chi_{Ran}=32\%$) displaying a discontinuity of more than ten words. Further, the average magnitude of the LVC size change is only $\Delta L_{Ran} = (30\pm 10)$ words, a full order of magnitude smaller than in the normative cases. Therefore explosiveness characterises normative acquisition as a genuine pattern of language learning. 

Is the explosive appearance of the LVC due to the acquisition of specific links or rather to specific words? In order to test this, we focus on the set of ``critical'' words, i.e.\ the single words whose addition allows for the sudden emergence of the LVC. We then compare features of these critical words with features of words already within the LVC at the time of its emergence. We test features like node-attributes (e.g.\ frequency, polysemy scores, etc.) and node degree. At a 95\% confidence level, no difference was found for any feature (sign test, $p-value = 0.007$). This lack of difference suggests that the emergence of the LVC is indeed due to higher-order link correlations rather than local topological features (such as degree) or psycholinguistic attributes. Hence, it is the global layout of links that ultimately drive the explosive appearance of the LVC. As shown also in Fig.\,\ref{fig1} (c-e), links crucial to the formation of the viable cluster might be acquired earlier (Fig.\,\ref{fig1} (c)) but the LVC might appear only later (Fig.\,\ref{fig1} (e)), after some key pathways completing the viable cluster are added to the network (Fig.\,\ref{fig1} (d)). 

The explosive emergence of the LVC has an interesting cognitive interpretation. Work in psycholinguistics suggests that frequency is the single most influential word feature affecting age of acquisition \cite{kuperman2012age} (mean Kendall $\tau$ $ \approx -0.47$ between frequency and AoA). We thus test whether the LVC growth can be reproduced through early acquisition of highly frequent words, with frequency counts gathered from Open Subtitles  \cite{barbaresi2014language}.
All simulations on the frequency-based ordering display an explosive emergence of an LVC ($\chi_{fre}=100\%$), however, the magnitude of the explosive transition is $\Delta L_{fre} = 280\pm 30$ words, which is only 2/3 of the normative one. At a confidence level of 95\%, the distribution of frequency-based LVC magnitude changes differs from the normative one (sign test, p-value = 0.01). The distribution of ages at which the LVC emerges in the frequency null model overlaps in 21\% of cases with the analogous normative one. However, we observe that the frequency null model differs from the normative one not only quantitatively (i.e.\ magnitude and appearance of explosiveness) but also qualitatively: the frequency null model displays a second explosive phase transition in LVC-size later in development, at around $10\pm 0.2$ years of age. This second transition might be due to the merging of different viable clusters, since we focused only on the largest viable cluster, rather than on viable clusters of non-trivial size. Further analysis reveals that the multiplex network has only one viable cluster, which suddenly expands through a second explosive transition in the frequency-based vocabulary growth model (but not in the normative AoA model). The above differences provide strong evidence that explosiveness in the mental lexicon is not an artefact of correlation of word frequency with language learning patterns.

We next test preferentially learning words with high degree in the multiplex network to see if the LVC emerges earlier than in normative acquisition. Learning higher degree words first makes more links available in the multiplex network. As we said above, it is links that drive the LVC emergence, hence we expect an earlier LVC appearance. The multidegree null model confirms this expectation and it displays a distribution of explosive transitions with average magnitude of $430\pm 30$ but happening almost two years earlier than in normative acquisition, around age $5.8 \pm 0.1$, cf.\ Fig.\,\ref{fig1}. The distribution of critical ages overlaps with the normative one only for 2\% of the time. We conclude that the degree acquisition is significantly different from the empirical case (mean Kendall $\tau \approx -0.31$ between multidegree and AoA).

Also word length influences lexical processing \cite{aitchison2012} and acquisition \cite{kuperman2012age}. Acquiring shorter words first leads to the sudden emergence of the LVC around age $6.6 \pm 0.6$, similarly to what happens for the polysemy curve. The LVC appears explosively with an initial size of $330\pm50$ words, a value lower than the normative one (mean Kendall $\tau \approx 0.24$ between word length and AoA). Differently from what happens with the polysemy curve, the growth of the LVC for shorter words is considerably faster compared to the normative case.

Another feature that can influence language acquisition is polysemy \cite{casas2016polysemy,sole2015ambiguity,sigman2002}, i.e. how many different definitions a word can have. We estimate word polysemy through polysemy scores \cite{casas2016polysemy}, including homonymy and also different meanings: the number of word definitions listed in the Wolfram dataset \textit{WordData} \cite{mathematica2016worddata}, which mostly coincides with WordNet. For a discussion about the caveats of using polysemy scores as we have defined above for quantifying polysemy we refer to SI Sect. 12. When words with higher polysemy scores are acquired earlier, we find the appearance of the LVC at around age $6.6 \pm 0.6$ years, with an average magnitude of $470\pm60$ words, close to the normative one. The distribution of critical ages at which the LVC emerges in the polysemy null model displays the highest overlap (35\%) with the analogous distribution from the normative case across all the null models we tested. Despite polysemy scores displaying a smaller correlation with the age of acquisition (mean Kendall $\tau \approx -0.26$) when compared to frequency or multidegree, it actually provides the highest overlap in terms of age at which the LVC emerges. This indicates that polysemy might play a role in driving the LVC emergence. 

Another attribute that could impact language development is concreteness, i.e.\ how tangible a given concept according to human judgements \cite{brysbaert2014concreteness,hanley2013concreteness}. Experimental research has shown that children tend to learn words earlier if a word is rated higher on concreteness \cite{kuperman2012age,brysbaert2014concreteness,piaget2000piaget,aitchison2012}. In order to test how concreteness can influence the LVC evolution, we develop a partial reshuffling null model (cf.\ Methods) where the  topology of words is fixed but node attributes are reshuffled at random. Partial reshuffling destroys the correlations between word features and the network topology, such that we can quantify the role of the relational structure in the absence of correlation with word features. Partial reshuffling gives rise to LVCs of the same size but containing words that are less concrete and less polysemous than in normative acquisition, cf.\ Fig.\ 2 (b). Partial reshuffling of word frequency leads to a gap in frequency of similar size as we see for concreteness (cf.\ SI Sect.\ 9). The gap in polysemy scores between the empirical and the reshuffled LVCs is five times larger than the analogous concreteness gap, suggesting that polysemy has a greater influence than concreteness over the emergence of the LVC. We also notice a peak in polysemy scores: the ``backbone'' of the LVC (i.e.\ the LVC emerging around 8 yr) is composed of significantly more polysemous words compared to the LVC at age 20 (cf.\ Fig.\ 2 (b), sign test, p-value = 0.001 $<$ 0.05). This early peak is absent in the partial reshuffling null model for polysemy scores. Furthermore, frequency (cf.\ SI Sect.\ 9) and concreteness do not display peaks early on after the LVC emergence. Such an early richness in high-polysemy words further indicates the idea that polysemy strongly influences the emergence of the LVC.

Even though potentially causing ambiguity in communication, polysemy is a universal property of all languages \cite{aitchison2012,sole2015ambiguity}. Conventionally when constructing semantic networks\cite{stey2005large,sigman2002,aitchison2012} word senses and meanings can be represented by links and polysemic words can have links related to different semantic areas (e.g.\ ``character'' is linked to ``nature'' in the context of complexion but also to ``font'' in the context of typography). Randomly Reshuffling word labels for all the neighbourhoods in the network evidently disrupts semantic relationships, thus destroying polysemy. We call this reshuffling ``full'' as it preserves the structure of local connections in the layers while fully destroying both intra-layer correlations at the endpoints of links and inter-layer correlations of words. We use full reshuffling as a null model (see Methods and SI) for testing how important polysemy is in determining the presence of the LVC. We fully reshuffle 2025 high-polysemy words (i.e.\ the words making up the heavy tail of the polysemy distribution) and compute the LVC size in the resulting reshuffled multiplex networks. Results are compared against a reference case in which the same number of low-polysemy words are fully reshuffled.
No viable cluster emerges on the multiplex networks with fully reshuffled high-polysemy words, while the LVC only shrinks by roughly $13\%$ in case of fully reshuffling low-polysemy words. We conclude that correlations between network structure and polysemy scores are indeed necessary in determining the presence of the LVC. 

The above results indicate that polysemy does increase lexicon navigability by ultimately giving rise to the LVC, i.e.\ a relatively small cluster of words that is fully navigable under both semantic, taxonomic, and phonological relationships in the mental lexicon. Such view is in agreement with previous works\cite{sole2015ambiguity,sigman2002,i2001small}, which point out how polysemy provides long-range connections in the lexicon which can increase navigability through different word clusters on semantic single-layer networks \cite{sigman2002}. 

\subsection*{Psycholinguistic characterisation of the Largest Viable Cluster (LVC)}

Next, we explore the impact of the presence of the LVC on cognitive aspects of language such as word processing. Our aim is to explore if words belonging to the empirical LVC (LVC-in) are processed differently than those words not in the LVC (LVC-out), more from a language use perspective rather than a developmental one (which was analysed with the previous null models). Hence, we turn to large-scale datasets of node attributes (see Tab.~1 and Methods). We find (cf.\ Tab.~1) that words in the largest viable cluster (i) are more frequent in the Open Subtitles dataset\cite{barbaresi2014language}, (ii) acquired earlier according to AoA reports \cite{kuperman2012age}, (iii) quicker to identify as words in lexical decision tasks \cite{keuleers2012british}, (iv) rated as more concrete concepts \cite{brysbaert2014concreteness} and 
thus more easily memorised \cite{hanley2013concreteness,binder2005distinct,brysbaert2014concreteness} and (v) represent more meanings in different semantic areas\cite{mathematica2016worddata,casas2016polysemy} when compared to LVC-out words.

In Fig.\ 3 (a-e), we report the cumulative probabilities of finding a word with a given feature less than a certain value for a set of particular node-level attribute within and outside of the LVC. The difference between LVC-in and LVC-out further indicates how different the words in the LVC are compared to LVC-out words. For instance, let us consider reaction times, which indicate how quickly people classify stimuli as words or nonwords in lexical decision tasks \cite{keuleers2012british}. The probability of finding at random an LVC-in word correctly identified in less than $500\,ms$ is $0.48$ while the same probability is less than half, $0.2$, for LVC-out words. Hence the LVC is rich in words identified more quickly. Analogous results hold for all the tested attributes.

Since LVC-in words have a higher degree compared to LVC-out words (see SI Sect.\ 3) and degree correlates with many of the psycholinguistic attributes used in our study, it is interesting to quantify to what extent the difference between LVC-in and LVC-out is due to correlations with degree. Results shown below the thick line, in the lower part of Tab.~1, suggest that the degree effect does not fully explain the observed psycholinguistic features of the LVC: a sign test indicates that all the median node-attributes of LVC-in words are higher than those of LVC-out words, at 95\% confidence level. Notice that the  comparison that does not account for degree is still important since one could easily argue that degree itself can be interpreted as a cognitive component that affects word processing \cite{stey2005large,vitevitch2012complex}.

\begin{table}[htbp!]
\footnotesize
\centering
\caption{Average node attributes for words within the LVC and within the largest connected component (LCC) for each individual layer. All the values are medians, except for heavy-tail distributions such as the frequency and polysemy ones, where the arithmetic mean was used instead. All the values are sample-size corrected via Monte Carlo sampling. The last five rows refer to degree-corrected samplings, where the sampled LVC-out words have the same degree of the sampled LVC-in words. 
Error bars are reported in parentheses for brevity: 3.93(3) means $3.93 \pm 0.03$.}
\begin{tabular}{|l|c|c|c|c|c|c|c|c|}
\hline
Node Attributes & LVC-in & LVC-out & Asso. LCC-in & Syno. LCC-in & Hyp. LCC-in & Phon. LCC-in & LCC Int. & LVC Rew.\\
\hline
\hline
Age of Acquisition [ys] & 6.43(2) & 9.4(1) & 8.5(1) & 8.8(1) & 9.0(1) & 7.8(1) & 7.4(1) & 7.3(1)\\
\hline
Concreteness [rating] & 3.93(3) & 2.83(4) & 3.63(4) & 3.35(5) & 3.45(5) & 3.87(2) & 3.72(2) & 3.71(3)\\
\hline
Reaction Times [ms] & 552(1) & 600(3) & 579(1) & 581(2) & 588(2) & 581(2) & 576(1) & 569(1)\\
\hline
Log Frequency [Counts] & 3.40(1) & 2.57(1) & 2.86(1) & 2.85(1) & 2.79(1) & 2.95(1) & 3.20(1) & 3.30(1)\\
\hline
Polysemy [Meanings] & 9.7(2) & 3.6(2) & 4.9(1) & 5.6(1) & 4.6(2) & 5.8(1) & 7.6(1) & 8.2(1)\\
\hline
Word Length [Letters] & 4.43(3) & 6.95(3) & 6.35(3) & 6.29(3) & 6.58(3) & 4.89(2) & 4.76(2) & 4.85(1)\\
\hline
\hline
\thickhline
\hline
Degree Corrections & LVC-in & LVC-out & Asso. LCC-in & Syno. LCC-in & Hyp. LCC-in & Phon. LCC-in & LCC Int. & LVC Rew.\\
\hline
Age of Acquisition [ys] & 6.43(2) & 7.62(1) & 7.2(1) & 8.1(1) & 8.1(1) & 7.5(1) & 6.62(2) & 6.61(2)\\
\hline
Concreteness [rating]& 3.93(2) & 3.67(3) & 3.79(2) & 3.42(4) & 3.40(4) & 3.89(2) & 3.89(2) & 3.90(2) \\
\hline
Reaction Times [ms] & 552(1) & 565(1) & 559(1) & 570(2) & 566(2) & 575(3) & 556(1) & 555(1)\\
\hline
Log Frequency [Counts] & 3.40(2) & 2.86(2) & 3.26(1) & 3.21(1) & 3.21(1) & 3.30(2) & 3.32(1) & 3.36(1)\\
\hline
Polysemy [Meanings] & 9.7(2) & 5.48(2) & 6.8(1) & 8.0(1) & 7.7(1) & 6.4(2) & 8.5(1) & 8.7(1)\\
\hline
Word Length [Letters] & 4.43(3) & 6.15(3) & 5.88(2) & 5.84(5) & 6.16(5) & 4.54(2) & 4.49(2) & 4.52(2)\\
\hline
\end{tabular}
\end{table}

\begin{figure}[htbp!]
\centering
\includegraphics[width=14cm]{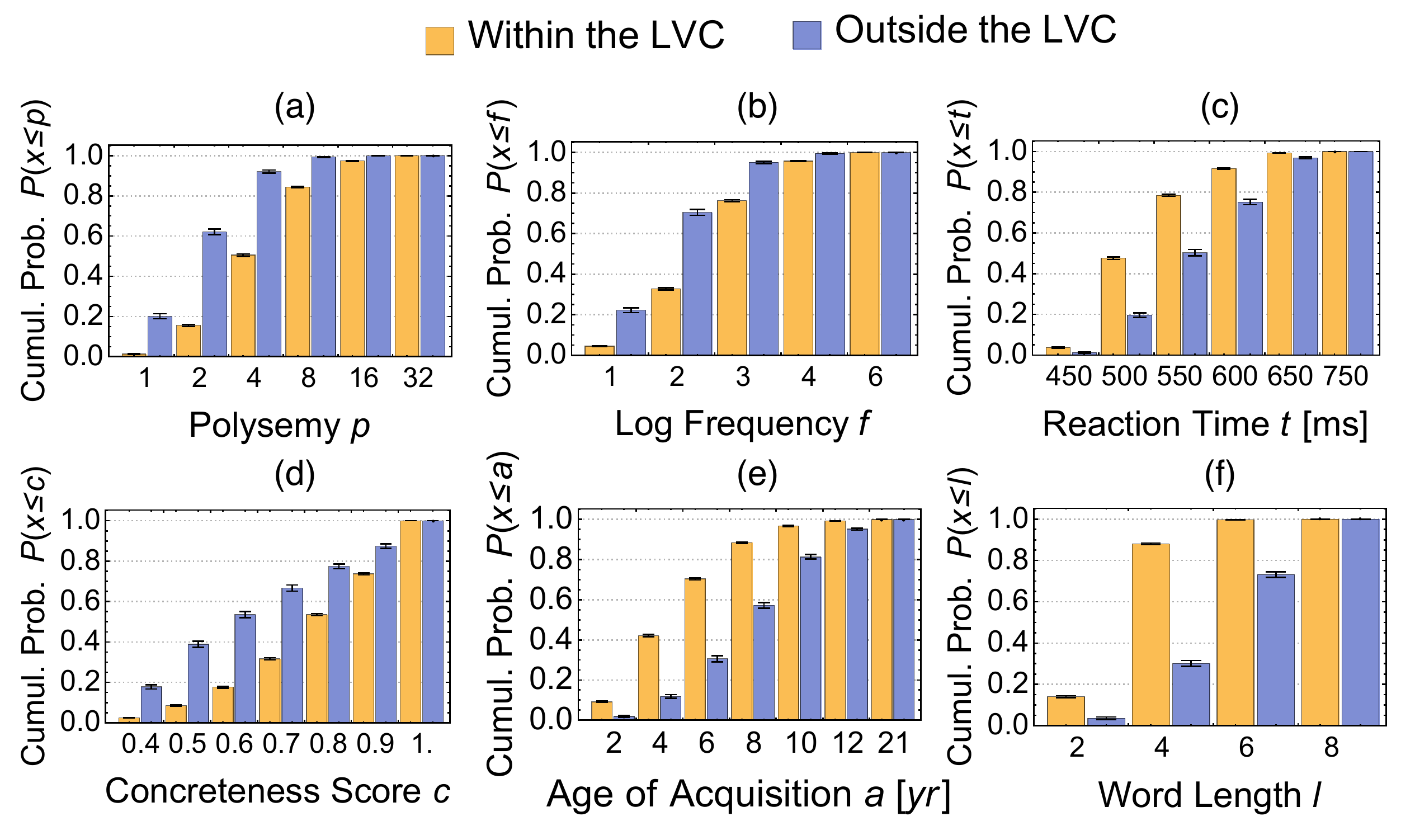}
\caption{ Cumulative probabilities of finding a word with a given feature less than a threshold T for LVC-in (orange boxes) and LVC-out (blue boxes). Concreteness scores are renormalised between 0 and 1 for easier binning. As an example, the probability of finding a low frequency word ($f \leq 10$) at random is 0.05 for LVC-in words but almost five times larger for LVC-out words.}
\label{fig3b}
\end{figure}

Tab.~1 also compares the statistics of the LVC against its single-layer counterparts, i.e.\ the largest connected components \cite{de2013mathematical} (LCC-In). We also consider multiplex alternatives to the LVC such as: the intersection across all layers of words in the LCC of each layer (LCC Int, cf.\ SI Sect.\ 8) and the LVC-in configuration models (LVC Rew.), which consist on average of 40\% more words. The empirical LVC consists of words with the most distinct linguistic features compared to the other tested sets of words, in terms of all tested node attributes. Even rewiring all links does not completely disrupt such distinctness (cf.\ LVC Rew.). These differences in linguistic attributes suggest that the LVC is a better measure of ``coreness'' for words in the mental lexicon than either the LCCs or their intersection, an idea we test further in the next section.

\subsection*{Robustness of the multiplex lexicon and LVC to cognitive impairments}
\begin{figure}[ht]
\centering
\includegraphics[width=15cm]{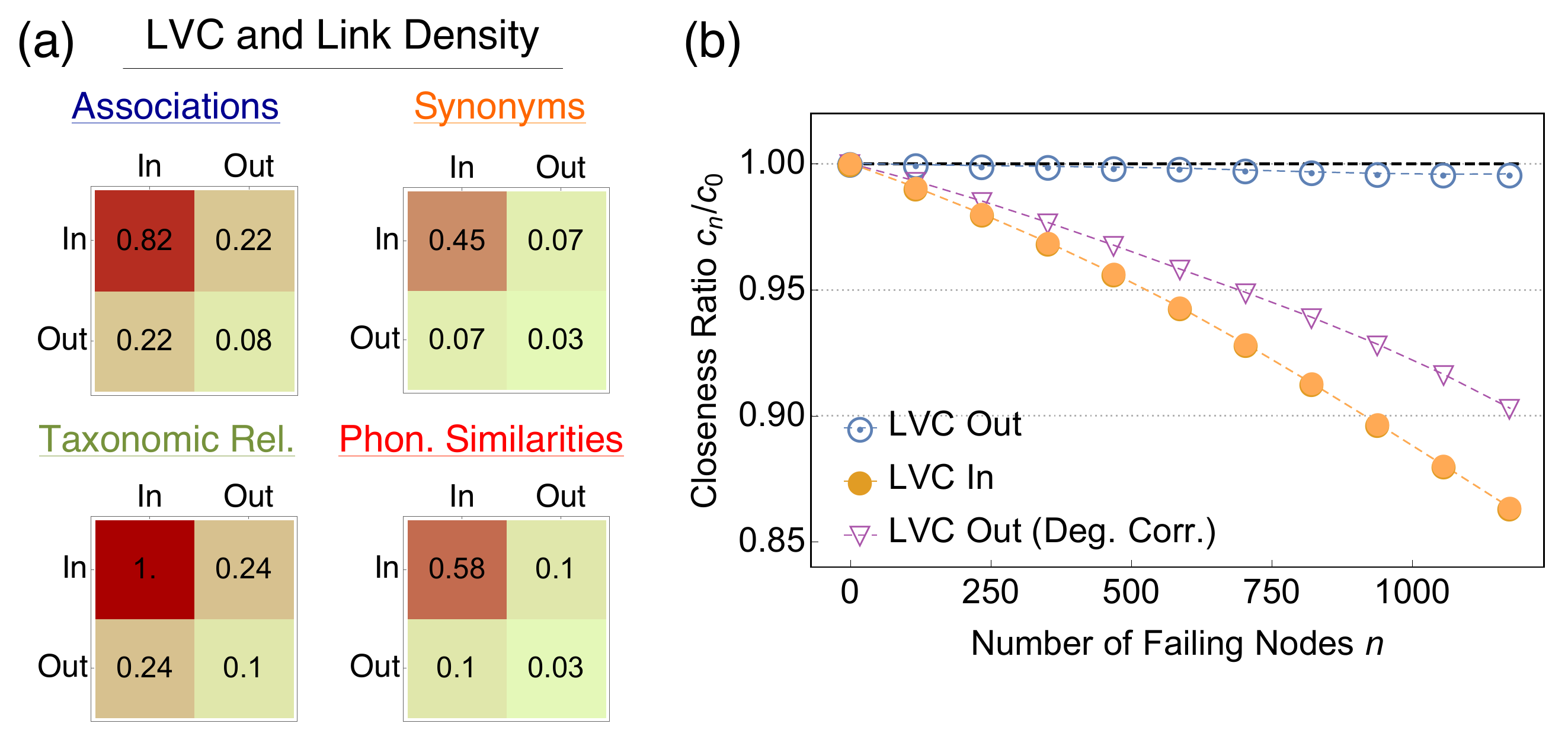}
\caption{(a) Normalised link densities across layers for couples of nodes either in the LVC (In), out of the LVC or on the boundary (one node in, one node out). Densities are normalised by the maximum value ($Lp_{In/In}$ for taxonomic relations) and colour coded (the higher the value, the more red the cell). (b) Resilience analysis with respect to random word failure, mimicking progressive aphasia in the mental lexicon. Words are targeted at random and then removed from the whole multiplex. In LVC-Out (Deg. Corr.) we remove words from outside the LVC but with the same degree as the words removed inside the LVC, thus correcting for a degree effect seen in the LVC which will also effect efficiency. As a measure of efficiency we use the median closeness of words in the network, providing the inverse of the average number of network hops necessary for reaching any word from any other one through the multiplex topology. Error margins represent standard deviations and they are about the size of the dots.}
\label{fig3}
\end{figure}

The LVC has been characterised as a set of higher degree words that differ in psycholinguistic features when compared to words located outside the LVC in our multiplex. This suggests that the higher degree, and cognitive correlations, of the LVC may be because the LVC is acting as a core for the mental lexicon. Let us denote the total number of links on a given layer as $L$ and the link density as $p$. As shown in Fig.\,\ref{fig3} (a), there are more links within the LVC ($Lp_{In/In}$) across all layers than outside of it ($Lp_{Out/Out}$) or at the interface of the LVC ($Lp_{In/Out}$). Further, across all individual layers the inequality $p_{In/In}>p_{In/Out}>p_{Out/Out}$ holds, denoting the presence of a core-periphery structure for the node partition $\{In,Out\}$ \cite{newman2012communities}. 

In order to better interpret both the coreness and cognitive impact of the LVC, we perform a resilience analysis of the MLR by means of numerical experiments. Random word failure provides a plausible toy model for progressive anomia\cite{laine2013anomia} driven by cognitive decline, where words become progressively non-accessible on all the lexicon levels without a clear trend \cite{laine2013anomia}. 

To simulate progressive anomia, we randomly remove LVC-in and LVC-out words in separate experiments. The maximum number of removed words is 1173, corresponding to the size of the LVC. As a proxy for robustness, we consider the average multiplex closeness centrality, which correlates with the average cognitive effort for identifying and retrieving words within the lexicon \cite{collins1969retrieval,sigman2002} and plays a prominent role in early word acquisition as well \cite{stella2016multiplex}. The results of this analysis are shown in Fig.\,\ref{fig3} (b).

We find that the multiplex representation is robust to random LVC-out word removal: removing almost 1170 LVC-out words only reduces average closeness, a measure previously linked to cognitive navigation \cite{vitevitch2012complex,stella2016multiplex,sigman2002,collins1975spreading}, to a level that is still within a 95\% confidence level of the original multiplex. Therefore failure of LVC-out words does not impact the cognitive effort in identifying and retrieving words within the lexicon. Instead, the multiplex lexicon is fragile to random LVC-in word removal: removing 50\% of words from the LVC leads to a decrease in closeness 20 times larger than the drop observed for LVC-out words. While considering random removal in both cases, it is true that in general LVC-in words have higher degree than LVC-out words, which might influence the robustness results from a technical perspective. The discrepancy in closeness degradation is only partly due to the higher degree of LVC-in words. Performing degree-corrected LVC-out word deletions still leads to less of a decrease in navigability as compared to LVC-in word deletion, as evident from Fig.\,\ref{fig3} (b). 

In summary, the multiplex lexicon is fragile to word failures of LVC-in words and robust to random failures of LVC-out words. This difference is a strong indicator that the LVC provides the necessary short-cuts for efficient navigation -- with high closeness and thus low cognitive effort -- of the mental lexical representation. It is worth remarking that the network's navigability is expected to increase in the presence of cores\cite{csermely2013structure,newman2012communities}, further supporting the interpretation that the LVC acts as a core of the multiplex structure. It has been conjectured that the mental lexicon has a core set of concepts \cite{barsalou2008grounded,solonchak2015lexicon,aitchison2012,picard2009hierarchies}; we show here how various cognitive metrics can be correlated with the LVC, suggesting that future work may benefit from considering the LVC as a quantification of lexical core structure.

\section*{Discussion}

Previous literature from psycholinguistics has conjectured the existence of a core set of words in the lexicon \cite{barsalou2008grounded,solonchak2015lexicon,picard2009hierarchies,aitchison2012}. Here, for the first time, we give large-scale quantitative evidence to support these conjectures.
In fact, we identify the largest viable cluster (LVC) of words which: (i) favours the emergence of connectivity allowing for navigation across all layers at once and (ii) acts as a core for the multiplex lexical representation.
Words within the LVC display distinct cognitive features, being (i) more frequent in usage \cite{barbaresi2014language}, (ii) 
learned earlier \cite{kuperman2012age}, (iii) more concrete \cite{brysbaert2014concreteness} and thus easily memorised \cite{aitchison2012,brysbaert2014concreteness} and activating perceptual regions of the brain \cite{binder2005distinct}, (iv) more context-dependent meanings \cite{casas2016polysemy,mathematica2016worddata} and (iv) more easily identified in lexical decision tasks \cite{keuleers2012british} and (v) of shorter length \cite{kuperman2012age} than words outside the LVC.  Remarkably, the explosive emergence of the LVC happens around 7 years of age, which is also a crucial stage for cognitive development in children. According to Piaget's theory of cognitive development\cite{piaget2000piaget}, age 7 is the onset of the \textit{concrete operational stage}, in which children develop more semantic and taxonomic relationships among concepts (e.g.\ recognising that their cat is a Siamese, that a Siamese is a type of cat and that a cat is an animal, thus drawing the conclusion that their cat is an animal among several). Experimental evidence \cite{ozcan2012developmental} has also shown that, in this developmental stage, children display an increased ability of mental planning and usage of context-dependent words in a connected discourse such as narratives \cite{ozcan2012developmental}.
Interestingly, age 7-8 is also the onset of the so-called orthographic stage for the cognitive model of reading acquisition by Frith\cite{frith1985beneath}. Around age 7-8 years, children start recognising a large number of words automatically and instantly access their meaning, matching words to an internal lexicon that they have built up in the previous years. As a result, reading becomes much faster, as documented in experimental setups\cite{aitchison2012}. Age 7-8 is found to be crucial for cognitive development also by the empirical work of Gentner and Toupin\cite{gentner1986systematicity}, who showed how at that age the analogical reasoning improved dramatically in children. The emergence of the lexical core represented by the LVC around age 7 might support analogical reasoning through the acquisition of more metaphorical relationships. Once in place, the lexical core may improve the ability to acquire and connect new abstract words based on analogy at later stages. All these findings can be interpreted in terms of an increased ability to navigate context-dependent meanings in the mental lexicon, which we quantitatively link to the explosive emergence of LVC core structure above. This indicates that the multiplex lexical network is a powerful representation of the mental lexicon: the network structure can indeed capture and translate well-documented mental processes driving cognitive development into quantifiable information. Notice that the current study does not test whether the LVC causes such changes but  quantifies for the first time a change in the multiplex network structure that agrees with well documented developmental shifts in language learning and processing. \textit{Ad hoc} longitudinal studies in children around age 7 are needed in order to better relate the LVC emergence with specific psycholinguistic tasks related to proficiency in memory and language use.

From a psycholinguistic perspective, in our robustness experiments one could point out that removal of LVC-in words might increase the overall degree similarity of the remaining words, thus impairing retrieval of similar forms due to retrieval and recall issues, such as lemma selection \cite{aitchison2012}. While this effect agrees with the impairment expressed by the decrease in closeness, this drop cannot be attributed exclusively to increases in the similarity of degrees among words, due to removal of high degree LVC-in words. In fact, when we remove words with the same degrees both in the LVC and outside of it, closeness drops significantly more when removing LVC-in words. This strongly suggests that lemma selection issues due to degree similarities alone cannot explain the drop in closeness and the related “coreness” of concepts in the LVC.

One limitation of our current approach is that we do not consider lexical restructuring over time, i.e.\ the adults' representation of word relationships could be different compared to children's or adolescents'. Previous work on the phonological level \cite{storkel2002restructuring} showed partial differences in phonological neighbourhoods between pre-schoolers and pre-adolescents. However, we show that the LVC persists even when all connections are randomly rewired and the LVC still identifies relevant words, e.g.\ more frequent, more concrete, etc.\ suggesting that the role of the LVC may still hold even with restructuring. Link rewiring also allows consideration of the variance in word learning due to individual differences. Individual difference modelling may be especially important for quantification, diagnosing, explaining, and correcting various language learning and usage issues \cite{beckage2011small}.

Another limitation is that the network representation might not be exact, e.g.\ there might be spurious links in the empirical free association layer or mistaken phonetic transcriptions in the phonological layer. In order to address this issue, we randomly reshuffle 10\% of word labels, 2.5\% on each layer separately, and find that the largest viable clusters are 10\% smaller than the empirical LVC (t-test, $p-value=0.009$). However, the LVC after reshuffling exhibits analogous performance in the features discussed in Tab.~1 (sign test, $p-value=0.96$). Together with the random rewiring experiments, this is an indication that the LVC structure is robust to small perturbations due to errors in the annotation of links or word labels.

Core/periphery network organisation is commonly found in many real-world systems \cite{csermely2013structure,brede2009networks}, even though the definition of cores in multiplex networks remains an open challenge. We interpret the robustness experiments as quantitative indication that the LVC is acting as a core for the whole multiplex lexical network, increasing navigability in two ways. Within the LVC, words must be connected to each other, implying navigability from every word within the LVC across all individual layers. Outside of the LVC, connections to the viable cluster facilitate network navigation by making words closer to each other. Since closeness correlates with the cognitive effort in word processing \cite{vitevitch2012complex,collins1975spreading,collins1969retrieval,sigman2002}, the LVC can be considered as facilitating mental navigation through pathways of the mental lexicon. This quantitative result is in agreement with previous conjectures about multiple meanings facilitating mental navigation of words \cite{sigman2002,i2001small,sole2015ambiguity}. Additionally, our results also indicate that the LVC acts as a multiplex core. The core is robust to node failure due to densely entwined links and connections which allow for navigation even in cases where words become inaccessible, as in cognitive disorders like progressive anomia \cite{laine2013anomia}.
It is worth remarking that we identify such a core with the largest LVC as no other non-trivial viable cluster exists in the multilayer lexical representation.

Indeed, identifying a core in the mental lexicon provides quantitative evidence supporting previous claims\cite{barsalou2008grounded,solonchak2015lexicon} about the existence of a core of highly frequent and concrete words in the lexicon that facilitates mental navigation and thus word retrieval in speech production experiments \cite{barsalou2008grounded,solonchak2015lexicon,hanley2013concreteness}. Alongside the cognitive perspective, interpreting the LVC as a lexicon core provides support for further previous findings about the presence of a ``kernel lexicon'' in language \cite{dorogovtsev2001language,i2001small,picard2009hierarchies}, a set of a few thousand words which constitute almost 80\% of all written text \cite{aitchison2012} and can define every other word in language \cite{picard2009hierarchies}. Previous works on semantic \cite{dorogovtsev2001language,i2001small}, taxonomic \cite{picard2009hierarchies} and phonological \cite{vitevitch2012complex,siew2013community} single-layer networks identified a kernel lexicon for the English language with roughly 5000 words which has not changed in size during the evolution of languages. This kernel lexicon was identified with the largest connected component of the English phonological network \cite{siew2013community}. The LVC we present here is: (i) a subset of the phonological largest connected component and (ii) it also persists across semantic and taxonomic aspects of language. Hence, the LVC represents a further refinement of the kernel lexicon that (i) is rich in polysemous words, (ii) facilitates mental navigation and (iii) is robust to rewiring or cognitive degradation. These three features suggest an interpretation of the LVC as a linguistic core of tightly interconnected concepts facilitating mental navigation through key words.  

While the framework presented here has been applied only for the English language, comparison with other languages and linguistic representations to assess how universal the LVC core is remains an exciting challenge for future experimental and theoretical work.

\section*{Methods}

\subsection*{Dataset and cognitive interpretation}

The datasets used in this work come from different sources and thus the resulting multiplex network representation is based on independent studies. For the MLR we construct four layers that model semantic, taxonomic, and phonological relationships. We further distinguish semantic relationships in free associations and synonyms. For free associations, e.g.\ ``A reminds one of B'', we used the Edinburgh Associative Thesaurus \cite{coltheart1981mrc}. For both, taxonomic relations (e.g.\ ``A is a type of B'') and synonyms (e.g.\ ``A also means B'') we used WordData \cite{mathematica2016worddata} from Wolfram Research, which mostly coincides with WordNet 3.0 \cite{miller1995wordnet}. For phonological similarities we used the same dataset analysed in \cite{stella2015patterns} based on WordNet 3.0 \cite{miller1995wordnet}. We treat every layer as undirected and unweighted. Words in the multiplex representation are required to be connected on at least one layer.

Free associations indicate similarities within semantic memory, i.e.\ when given a cue word ``house'', human participants respond with words that remind them of ``house'', for example ``bed'' or ``home''. Networks of free associations play a prominent role in capturing word acquisition in toddlers \cite{hills2009longitudinal,stella2016multiplex} and also word identification \cite{de2016large,collins1975spreading}. Networks of synonyms are also found to play a role in lexical processing \cite{sigman2002,stey2005large,aitchison2012,baronchelli2013}. The hierarchy provided by taxonomic relationships deeply affects both word learning and word processing \cite{collins1969retrieval,sigman2002,aitchison2012,baronchelli2013}. Phonological networks provide insights about the competition of similar sounding words for confusability in word identification tasks \cite{vitevitch2008can,vitevitch2012complex,stella2015patterns}. 

For the linguistic attributes we combine several different sources. We source word frequency from OpenSubtitles \cite{barbaresi2014language}, a dataset of movie subtitles whose word frequencies were found to be superior to frequencies from classical sources in explaining variance in the analysis of reaction times from lexical decision experiments \cite{barbaresi2014language,keuleers2012british}.
Concretess scores \cite{brysbaert2014concreteness} and age of acquisitions ratings \cite{kuperman2012age} were gathered from Amazon Turk experiments, allowing for large-scale data collection and confirmation of previous findings based on small-scale experiments \cite{brysbaert2014concreteness,kuperman2012age}. Concreteness ratings indicate how individual concepts are rated as abstract (on a scale of 1 - ``abstract'' to 5 - ``concrete'')\cite{brysbaert2014concreteness}. Polysemy scores were quantified as the number of different definitions for a given word in WordData from Wolfram Research which coincides with WordNet \cite{mathematica2016worddata}.
Reaction times were obtained from the British Lexicon Project \cite{keuleers2012british} and indicate the response time in milliseconds for the identification of individual words were compared against non-words.

\subsection*{Smearing normative acquisition}

Smearing is a technique used in statistics for generalisation of data samples \cite{fayyad1996advances}. We smear the age of acquisition data from Kuperman et al.\cite{kuperman2012age}, where the average age of acquisition $a_i$ and standard deviation $\sigma_a(i)$ around each word are provided, e.g.\ {$a_{aim}=6.72\,yrs,\sigma_a(aim)=2.11\,yrs$}. In our case, smearing consists of sampling possible age of acquisitions for word $i$ from a Gaussian distribution $\mathcal{N}[a_i,\sigma_a(i)]$ rather than considering only the average value. Sampling independently an age of acquisition for each word in the dataset, we can build multiple artificial acquisition rankings from empirical data. Hence, smearing enables our analysis to account for not only the average ages of acquisition of words but also for their variability across individuals, thus adding robustness against individual variability to our results.

\subsection*{Lexicon growth experiments}

We simulate lexicon growth over time $t(n)$ by considering subgraphs of the multiplex lexicon where the first $n \leq 8531$ words in a given ranking $r$ are considered. 8531 is the total number of words in our network. Rankings indicate the way words are acquired in the lexicon over time and can be based on word features or age of acquisition reports. The rankings we use are based on: (i) smeared age of acquisition \cite{kuperman2012age}, (ii) frequency\cite{barbaresi2014language,kuperman2012age} (higher frequency words are learned earlier), (iii) multidegree\cite{de2013mathematical} (words with more links across all layers are learned earlier), and (iv) polysemy (words with more definitions are learned earlier). As a randomised null model, we consider random word rankings. When the first $n$ words in a ranking are considered, a subgraph of the multiplex lexicon with these words is built and its LVC is detected. By using the non-smeared age of acquisitions, we relate the number of learned words to the developmental stage in years $t(n)$, e.g.\ $n=1000$ corresponds to $t=5.5$ years. 

The size of the LVC $L(t)$ is then obtained as a function of developmental stage $t(n)$ for every specific type of ranking. Results for the smeared age of acquisitions and the random null model are averaged over an ensemble of $200$ iterations. Results for the frequency, degree, and polysemy orderings are averaged over $200$ iterations where words appearing in ties are reshuffled. Results are reported in Fig.\ 2.

Each iteration represents the evolution of the LVC size through the acquisition of an individual word. This acquisition trajectory may be related to different developmental stages. For every iteration, we detect the magnitude of the transition on the LVC size due to its appearance when adding words one by one to the network. We then compute the fraction $\chi$ of iterations presenting a discontinuity of more than 10 words entering into the LVC. We also compute the average magnitude of the explosive transition $\Delta L$. 

Comparisons of the empirical distributions of ages at which the LVC emerges considers the overlapping coefficient \cite{fayyad1996advances}, i.e.\ the overlap of two distributions normalised by the maximum overlap obtained when shifting the central moment of one of the distributions. An overlap of 100\% means that one distribution is fully contained in the other one. An overlap of 0\% means that the distributions have no overlap.

\subsection*{Robustness experiments}

We carried out robustness testing via word/node removal: individual words are removed at random across all layers. Closeness centrality is then measured by considering shortest paths across the whole multiplex network structure, i.e.\ also including jumps between layers. We consider closeness centrality as a measure for the spreading of information and the mental navigability of the lexicon \cite{collins1975spreading,i2001small,siew2013community}. In our case closeness is well defined, since even the deletion of the whole LVC leaves the multiplex network connected\cite{de2014navigability}. We consider a multiplex network as connected if it is possible to reach any pair of nodes by allowing for traversal along links on any layers. 

With reference to Fig.\ 3, we perform random attacks of words within the LVC (LVC-in) and outside of it (LVC-out). Since LVC-in words are more connected compared to words outside, we also perform degree corrected attacks: random words within the LVC and words of equivalent degree outside the LVC are removed. This degree correction (LVC-out - Deg.\ Corr.) allows for the attack of LVC-out words but reduces the number of links by the same amount as LVC-in attacks.

\subsection*{Data availability and Additional Information}

No new datasets were generated during the current study. The list of LVC-in and LVC-out words is available online at https://goo.gl/Dd9eC6. The authors declare no competing financial interests. Material requests should be addressed to the corresponding author.

\bibliography{sample}

\section*{Acknowledgements}

M.S. was supported by an EPSRC Doctoral Training Centre grant (EP/G03690X/1). M.D.D. acknowledges financial support from the MINECO (Spain) program ``Juan de la Cierva'' (IJCI-2014-20225).

\section*{Author contributions statement}

M.S., N.B., M.B. and M.D.D. conceived the experiments, M.S. overlapped and cleaned the data, M.S. performed the experiments, M.S., N.B., M.B. and M.D.D. analysed the results. All authors reviewed the manuscript. 

\end{document}